# Electron Lens as Beam-Beam Wire Compensator in HL-LHC

A. Valishev and G. Stancari, FNAL, Batavia IL 60510, USA
21 November 2013

Current wires are considered for compensation of long-range beam-beam interactions for the High Luminosity upgrade (HL-LHC) of the Large Hadron Collider at CERN [1]. In this note, we demonstrate the advantage of using Electron Lens for this purpose instead of a conventional current-bearing wire.

First, let us consider the required strength of a beam-beam wire compensator. In the simple case of round beam (or large transverse beam-beam separation), a test particle moving at a distance $r$ from the center of an opposing proton bunch of charge $N_p$, experiences a radial kick

$$\Delta p_r = \int e(E_r + \beta c B_\varphi)dt = eN_p c(1+\beta)\frac{e}{2\pi\varepsilon_0 c^2}\frac{1}{r},$$

where $e$ is the electron charge, $c$ is the speed of light, $\beta$ is the relativistic factor of the beam. We define the long-range kick strength for one encounter as

$$\mathcal{L}_1 = eN_p c(1+\beta)/2 \cong ecN_p = ec \times 2.2 \cdot 10^{11} = 10.5\ A\cdot m$$

Considering that in the ideal case of $N_{LR}$=18 long-range encounters on one side of the main IP, a single compensator would cancel the kick from all of them, the required strength of the compensator is

$$\mathcal{L}_{LR} = N_{LR}\mathcal{L}_1 = 18 \times \mathcal{L}_1 = 190\ A\cdot m$$

If a single wire of length $L$ with current $I$ is used for compensation, the kick is created through the magnetic field of the current:

$$\Delta p_r = \int e\beta c B_\varphi dt = (L \cdot I) \times \frac{e}{2\pi\varepsilon_0 c^2}\frac{1}{r}$$

Thus, a wire would need to carry $\mathcal{L}_{Wire} = 190\ A\cdot m$ over its length.

However, if one uses a low-energy electron beam in an Electron Lens for such compensation, the kick is created through the electrical field of the electron column as well as through the magnetic field of the current:

$$\Delta p_r = \int e(E_r + \beta c B_\varphi)dt = (L \cdot I)\frac{1 \pm \beta\beta_e}{\beta\beta_e} \times \frac{e}{2\pi\varepsilon_0 c^2}\frac{1}{r}$$

Here $\beta_e$ is the relative velocity of electrons in the electron lens, and the sign in the numerator is chosen based on the relative direction of propagation of the electron beam. Typically, the energy of electrons in an electron lens device is low (5 keV in the Tevatron Electron Lens), hence the current necessary for long-range compensation can be reduced by a factor $\frac{1+\beta\beta_e}{\beta\beta_e} = \frac{1+0.2}{0.2} = 6$, which gives for the HL-LHC parameters the strength

$$\mathcal{L}_{E-Lens} = \frac{\mathcal{L}_{Wire}}{6} = 32\ A\cdot m$$

Tevatron Electron Lenses were capable of transferring the current of up to $2\ A$ over the length of $2.5\ m$ [2]. Since the beam size requirements for the long-range compensator are much relaxed compared to the head-on application of TEL's, a device transporting the electron current of $10\ A$ over $3\ m$ is technically feasible.

Additionally, an electron lens based beam-beam compensator offers the capability to easily adjust the compensation strength for different bunches in a train (in particular, for the so-called pacman bunches) because the current pulse amplitude can be varied on a bunch-by-bunch basis.

*Fermi Research Alliance, LLC operates Fermilab under contract No. DE-AC02-07CH11359 with the U.S. Department of Energy.
This work was partially supported by the US LHC Accelerator Research Program (LARP).